\documentclass[aps,pra,reprint]{revtex4-1}
\usepackage{epsfig}
\usepackage{graphicx}
\usepackage{amsmath}
\usepackage{amsfonts}
\usepackage{amssymb}
\usepackage{graphicx}
 \usepackage{wrapfig}
\usepackage[cp1251]{inputenc}
\usepackage{epsfig}
\usepackage{graphicx}
\usepackage{amsmath}
\usepackage{amsfonts}
\usepackage{amssymb}
\usepackage{graphicx}
 \usepackage{wrapfig}
\usepackage{textcomp}
\usepackage{pgf,pgfarrows,pgfnodes,pgfautomata,pgfheaps,pgfshade}
\usepackage{amsmath,amssymb}
\usepackage[config]{caption,subfig}
\usepackage[colorlinks=true, linkcolor=blue]{hyperref}
\usepackage{marginnote}  
\newcommand{{\sign}}{\rm sign}
\newcommand{{\const}}{\rm const}

\begin{document}
\title{Synchronization of qubit ensemble under optimized $\pi$-pulse driving}
\author{Sergey V. Remizov$^{1,2}$}
\email{sremizov@gmail.com}
\author{Dmitriy S. Shapiro$^{1,2}$} \author{Alexey N. Rubtsov$^{1,3,4}$}
\affiliation{$^1$Center for Fundamental and Applied Research, N. L. Dukhov All-Russia Research Institute of Automatics,  127055 Moscow, Russia}
\affiliation{$^2$V. A. Kotel'nikov Institute of Radio Engineering and Electronics, Russian Academy of Sciences, 125009 Moscow, Russia}
\affiliation{$^3$Russian Quantum Center,   Skolkovo, 143025 Moscow Region, Russia}
\affiliation{$^4$Department of Physics, Moscow State University, 119991 Moscow, Russia}

\begin{abstract}
We propose technique of simultaneous excitation of disordered  qubits  providing an effective suppression of  inhomogeneous broadening in their spectral density.  The technique is based on applying of optimally chosen  $\pi$-pulse with non-rectangular smooth shape.   We study  excitation dynamics of an off-resonant qubit subjected to strong classical electromagnetic  driving field with the fast reference frequency and slow envelope. Within this solution we optimize the envelope to achieve a preassigned  accuracy in qubit synchronization.
\end{abstract}

\maketitle

\section{Introduction}

The investigation of qubit ensembles reveals analogies  with quantum optics effects  \cite{Blais, YouNori,Astafiev}   and possibilities for construction of quantum computers and simulators  \cite{DiCarlo, MSS, Nation,  Clarke}.  Solid state realizations of qubit ensembles are superconducting Josephson circuits  \cite{MSS, Orlando,mooij},  nitrogen-vacancy (NV) centers in diamond samples \cite{nv-centers-0,nv-centers, nv-centers-1}, or nuclear and electron spins realized as $^{31}$P donors in $^{28}$Si crystals \cite{Morton} and Cr$^{3+}$ spins in Al$_2$O$_3$ \cite{Schuster}.  The  coupling of qubit ensembles  with a superconducting  microwave resonators results in the formation of  sub-wavelength quantum metamaterials \cite{Macha,Rakhmanov, Fistul,ZKF,SMRU}.  The long-range interaction through a photon mode  results in the formation of  collective qubit  states in such metamaterials \cite{Brandes,Zou}, as Dicke model describes. One of crucial distinctions of artificial qubits from natural atoms is that their excitation energies are in many cases tunable  \textit{in situ}  by external magnetic fields. Beside of the tunability, another  property  is a disorder in excitation frequencies and, as a consequence, inhomogeneous broadening of the density of states in qubit ensembles. This is related to  fundamental mechanisms  such as exponential dependence of excitation energy on Josephson and charging energies in superconducting qubits or spatial  fluctuations of  background magnetic moments \cite{Stanwix} in systems with NV-centers.

 Disordered spectrum  of collective modes offers    multimode quantum memory,  where information about  photon state is encoded  as  a tunable collective qubit mode \cite{Wesenberg0,Wesenberg}. The storage and retrieval  protocols were proposed in Refs.   \cite{Moelmer, Grezes, Wu} and based on spin-refocusing techniques \cite{spinref-1,spinref-2} or successive magnetic field gradients. 
In the context of quantum memory  
the unavoidable spectral broadening in qubit excitation frequencies provides multimode performance from  one side, but from the other side this is one of limiting factors affecting coherence times. Therefore, the development of  techniques of effective suppression of the disorder in qubit frequencies and synchronization of their dynamics  is an important problem.  For instance, one of the options is   atomic frequency comb (AFC) technique which could be applied to rare-earth-metal-ion qubit ensembles  \cite{ASRG}. This method is based on frequency-selective optical pumping and subsequent transitions to metastable auxiliary hyperfine states.
 Another way of solution of this problem was demonstrated in    Ref.   \cite{nv-centers} as 'cavity protection' effect in NV-centers.
   The effect 
 is related to a decreasing of  a relaxation rate of collective qubit modes which is proportional to the spectral broadening.

  Our research is  inspired by  one of the key  ideas of Ref.   \cite{nv-centers}:
 the  succession of  microwave rectangular pulses  can serve as  efficient method for  excitation of disordered NV-centers from the ground to the excited state.
 In our paper we study the possibility of simultaneous qubit excitation by a \textit{single} non-rectangular $\pi$-pulse, rather than the  sequence mentioned above.
  We observe that the optimized  non-rectangular shape of   $\pi$-pulse provides an efficient tool for suppression of the  disorder effects as well. It  allows to excite qubits  within a wide detuning range with almost 100\%  probability.  In contrast to AFC methods, this technique does not require using of auxiliary levels transitions.

 We assume that the $\pi$-pulse is realized as electromagnetic signal  $f(t)e^{-i\omega t}$ of a carrying frequency $\omega$ being almost in resonance with the qubit
excitation frequency.
 In our solution we perform the optimization of the envelope shape $f(t)$ in an experimentally relevant class of smooth
 functions, which guarantees that higher energy levels of a qubit are not  affected.
  We expect that this technique can be applied to disordered systems with strong qubit-cavity couplings like NV-centers or superconducting  metamaterials, as well as to the atomic clock devices \cite{Hodges} as a tool for the preparation of a particular atomic state.

\section{Definitions}
We address to the possibility of simultaneous excitation of disordered qubit ensemble coupled to  photon transmission line being the source of the driving. Qubits are assumed to be non-interacting with each other and long lived in comparison to $\pi$-pulse duration time $\tau\ll\tau_\varphi$.
The absence of qubit-qubit interactions means that we can study  dynamics of a single off-resonant driven qubit. We fix carrying frequency $\omega$ and assume that the qubit energy $\epsilon$ can be varied reflecting the spectral broadening. Neglecting the qubit decoherence we  solve  the Schr\" odinger equation only $i\partial_t| \psi\rangle=(H_{q}+H_{ext})|\psi\rangle$, where unperturbed qubit Hamiltonian is  $H_q=\epsilon(\sigma_0 + \sigma_z)/2$ and the external driving is $H_{ext}=(f(t)e^{-i\omega t}\sigma_++f^*(t)e^{i\omega t}\sigma_-)/2$.
We define wave function of the qubit state in   $\omega$-rotating frame  as
$$
|\psi(t)\rangle
=
\begin{pmatrix}
\alpha(t)e^{-i(\omega +\delta/2)t} 
\\[.5em] 
\beta(t)e^{- i \delta t/2}\end{pmatrix},$$
where the  detuning frequency is $\delta=\epsilon-\omega$. The  Hamiltonian $H$ of the driven  qubit in this rotating frame reads
\begin{equation}
H=\frac{1}{2}\begin{pmatrix}
\delta && f(t)\\[.5em]
 f^*(t) && -\delta
\end{pmatrix}\label{h}
\end{equation}
 We assume that the qubit is close to cavity resonance $\epsilon\approx\omega$ and consider the evolution of the qubit wave function within the time interval $0<t<\tau$ starting from the ground state  $|\psi(0)\rangle=|g\rangle$ at the initial moment of time $t=0$.
   The $\pi$-pulse time $\tau$ is  considered   as a fixed value. At this point we define frequency
   $$
  \Omega=\frac{\pi}{\tau},
   $$
 which is the main scale in our consideration along with the detuning $\delta$.
The frequency $\Omega$ has the  transparent physical meaning: this is frequency of Rabi oscillations of the resonant qubit with $\delta=0$ under
 the constant driving amplitude given by $F_0(t)=\Omega e^{-i\omega t}$. Hence, time $\tau $  is the half of the Rabi period associated with rectangular  $\pi$-pulse $F_0(0<t<\tau)$   exciting the  system from  $|g\rangle$ to $|e\rangle$. Non-zero detuning, related to  inhomogeneous broadening or spread in qubit frequencies, does not allow to achieve full qubit excitation if envelope shape $f(t)$ is constant. In the following consideration we modify $f(t)$ at the time interval $0<t<\tau$ to more complicated non-rectangular shape $f(t)\neq \const$ to achieve higher efficiency in  near-to-resonance qubit excitation.

Schr\" odinger equation with the Hamiltonian (\ref{h}) allows analytical solutions   only in several particular cases. The basic one is the constant driving amplitude  $f(t)=f=\const$ and arbitrary detuning $\delta$ which corresponds to damped  Rabi oscillations of the frequency $\Omega_R=\sqrt{f^2+\delta^2}$. In this case the evolution of the wave function being in the ground state  at $t=0$ reads
 \begin{equation}
|\psi(t)\rangle=\begin{pmatrix}
-\frac{if}{\Omega_R}\sin \Omega_R t/2 \\ \\
\cos\Omega_R t/2 + \frac{i\delta}{\Omega_R}\sin \Omega_R t/2
\end{pmatrix}.\label{rabi0}
\end{equation}
One can see that detuning reduces the maximum of excitation probability. This effect results in impossibility of  synchronization of qubit ensemble by the rectangular shape of the  driving envelope.

Our further consideration is based on another   exact solution, which holds for   on-resonance driving regime  $\delta=0$  and arbitrary  real valued  $f(t)$. The time evolution of the ground state within this solution reads
 \begin{equation}
|\psi(t)\rangle=\begin{pmatrix}
\displaystyle
-i  \sin\frac{\varphi(t)}{2}
\\[1em]
\displaystyle
\cos\frac{\varphi(t)}{2}
\end{pmatrix},\label{rabi0}
\end{equation}
where the phase $\varphi(t)$ is given by the time integral
 $$
 \varphi(t)=
    \int\limits_0^t f(t_{1}) dt_{1}.
 $$
The  $\pi$-pulse condition, which is  the inversion of a qubit occupation number, for this resonant case  holds for
\begin{equation}
\varphi(\tau) = \pi.
\label{eq:pipulse}
\end{equation}
The last constraint (\ref{eq:pipulse}) provides the class of real valued functions $f(t)$ we are addressed to in the optimization procedures below.

\section{Perturbative solution}
As it was  mentioned above, the  exact solution is not known for arbitrary $f(t)$ and non-zero detuning $\delta\neq 0$. Hence, we develop a perturbation theory based on  treating the $\delta\sigma_z/2 $-terms in the Hamiltonian (\ref{h}) as small perturbation and considering the exact solution (\ref{rabi0}) at $\delta=0$   as the zero order approximation.
We end up with the following recursive equations forming the perturbation theory by $\delta$
\begin{align}
  i \dot \alpha^{(n)}(t)
  - &\,
  \frac{f(t)}{2} \beta^{(n)}(t)
  =
  \frac{\delta}{2}\alpha^{(n-1)}(t)\label{eq:abnb1}
  \\
  i \dot \beta^{(n)}(t)
  - &\,
  \frac{f(t)}{2} \alpha^{(n)}(t)
  = -
  \frac{\delta}{2}\beta^{(n-1)}(t)
  .\label{eq:abnb2}
\end{align}
The full solution reads
\begin{equation}
  \begin{pmatrix}
  \alpha(t)
  \\[1em]
  \beta(t)
  \end{pmatrix}
  =
  \begin{pmatrix}
  \displaystyle
  x(t) \cos\frac{\varphi(t)}{2}
  -i y(t) \sin\frac{\varphi(t)}{2}
  \\[1em]
  \displaystyle
  y(t) \cos\frac{\varphi(t)}{2}
  -ix(t) \sin\frac{\varphi(t)}{2}
  \end{pmatrix}
  . \label{eq:solab}
\end{equation}
Assuming that the qubit was in the ground state at the initial moment of time $|\psi(0)\rangle=|g\rangle$, the solution for $x(t)$ and $y(t)$  is given by  $\delta^n$ series with nested integrals as the coefficients
\begin{align} \label{eq:x}
    x(t)
  = &\,
  -
  \frac{\delta}{2} \int \limits_0^t dt_{1} \sin \varphi(t_1)
  \\ &\,
  +
  \frac{\delta^2 i}{4}
  \int\limits_0^{t} dt_{1}
  \int\limits_0^{t_{1}}  dt_{2}\,
  \sin \left[
    \varphi(t_2)
    -
    \varphi(t_1)
  \right]
  +
  \dots \nonumber \\ \label{eq:y}
  y(t)
  = &\,
  1
  +
  \frac{\delta i}{2} \int \limits_0^t dt_{1}
  \cos \varphi(t_1)
  \\ &\,
  -
  \frac{\delta^2}{4}
  \int\limits_0^{t} dt_{1}
  \int\limits_0^{t_{1}} dt_{2}\,
  \cos \left[
    \varphi(t_2)
    -
    \varphi(t_1)
    \right]
  +
  \dots\nonumber
\end{align}
These are general  equations of the perturbation theory we use to find optimal shape of a $\pi$-pulse   $f(t)$   providing simultaneous qubit excitations.
Assuming envelope of the driving as smooth function at $t=0$ and $t = \tau$ we model $f(t)$ as  superposition of finite number of $N$ sine functions,  where $\Omega$ comes as a factor both in sine arguments and the driving amplitude
\begin{equation}
   f(t)=\Omega\sum\limits_{n=1}^N k_{2n-1}\sin( n-1/2)  \Omega t.
   \label{eq:phisum}
\end{equation}

According to (\ref{eq:solab}) the wave function after the $\pi$-pulse takes the following form
 \begin{equation}
  \label{eq:wf}
  \begin{pmatrix}
  \alpha(\tau)
  \\[1em]
  \beta(\tau)
  \end{pmatrix}=\begin{pmatrix}
  -iy(\tau)
  \\[1em]
  -ix(\tau)
  \end{pmatrix}.
  \end{equation} 
  We optimize numerically a finite set of coefficients $k_{2n -1}$ and require that ground state amplitude at $t=\tau$ is zero up to $\delta^N$ in $\delta$-expansion, i.e. $\beta(\tau)=O(\delta^N)$.  It is possible if two conditions are fulfilled: (i)  resonant qubit is excited, i.e. $\varphi(\tau)=\pi$, expressed in terms of $k_{2n-1}$ as
 \begin{equation}
  \label{eq:system:phi}
  4\sum\limits_{n=1}^N \frac{k_{2n-1}}{2n- 1} = \pi,
  \end{equation}
and (ii)  off-resonant qubit excitation almost does not depend on detuning, i.e. $x(\tau) = O(\delta^N)$ in (\ref{eq:wf}). The requirement $x(\tau) = O(\delta^N)$ can be  reduced to the  set of $N-1$ equations corresponding to vanishing  of $\delta^n$  terms  at $n\leq N-1$ in  the    expansion (\ref{eq:x}), if we represent it as $x(\tau,\delta)=\sum\limits_{n=1} c_n \delta^n$, where
\begin{align}
  \label{eq:system:sin1}
  c_1=\int \limits_0^{\tau} dt_{1} \sin \varphi(t_1)
  =
  0
  \\
  \label{eq:system:sin2}
  c_2=\int\limits_0^{\tau} dt_{1}
  \int\limits_0^{t_{1}}  dt_{2}\,
  \sin \left[
    \varphi(t_2)
    -
    \varphi(t_1)
  \right]
  =
  0
  \\ \nonumber
  \dots
\end{align}

To summarize, our perturbative solution consists of finite system of $N$ equations (\ref{eq:system:phi}, \ref{eq:system:sin1}, \ref{eq:system:sin2}, ...) providing smooth solution for $f(t)$. The  precision order of the technique is given by $N$ and  ensures that the excitation probability $n_\uparrow(\tau, \delta)$ of an off-resonant qubit is close to unity up to small   correction, namely $n_\uparrow(\tau, \delta)=1-|x(\tau,\delta)|^2$. This correction is nothing but the precision of the technique and is given by the probability of the ground state qubit occupation $n_0(\tau, \delta)\equiv|x(\tau,\delta)|^2$.

\section{Results}
\subsection{ $\pi$-pulse optimization  scheme at $N=3$  }
\begin{figure}[!ht]
  \includegraphics[width=\linewidth]{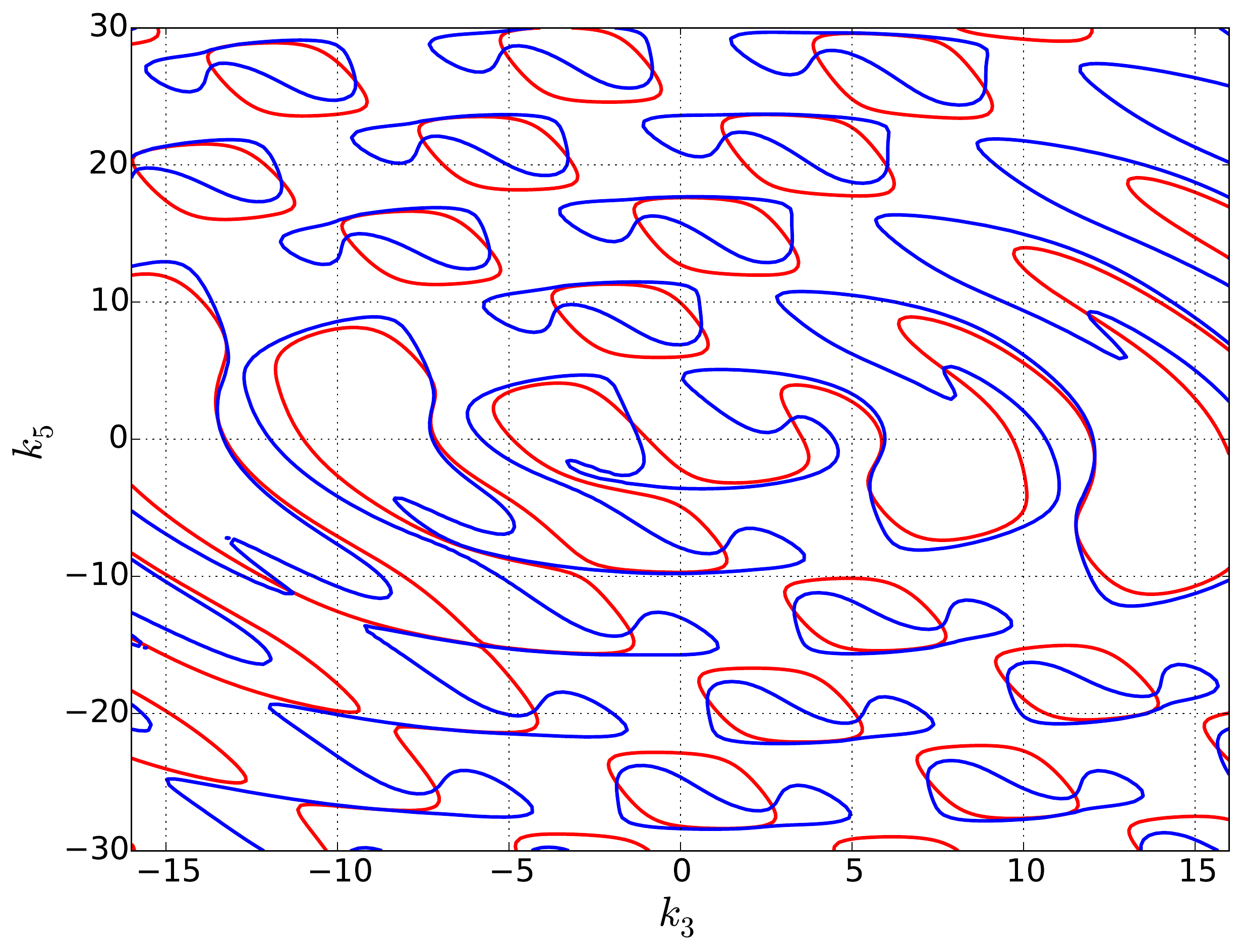}
  \caption{Curves in $(k_3, k_5)$ plane formed by optimal values of sine amplitudes. Red curve corresponds to quadratic dependence of ground state amplitude on detuning. Crossing points of red and blue curves correspond to cubic dependence of ground state amplitude on detuning. Such a point nearest to zero is marked by green circle and studied in details. The value of $k_1$ is selected so that $\pi$-pulse is formed.}
  \label{fig:map}
\end{figure}

In this subsection we provide numerical results for the optimization of $\pi$-pulse constructed from  $N=3$ terms
$$
  f(t)
  =\Omega(
  k_1 \sin t\Omega/2
  +
  k_3 \sin 3 t\Omega/2
  +
  k_5 \sin 5 t\Omega/2)
  .
$$
We start from  the numerical solution for $k_3$ and $k_5$ using (\ref{eq:system:sin1},\ref{eq:system:sin2}) and after that we choose $k_1$ according to $\pi$-pulse  condition constraint (\ref{eq:system:phi}). In the fig.\ref{fig:map} we plot two sets of curves  in coordinates $(k_3, k_5)$: (i) red curves correspond to vanishing of the linear in $\delta$ term in $x(\tau)$, i.e. $c_1=0$, according to Eq. (\ref{eq:system:sin1}), hence, in this case the residual part for the ground state amplitude is $x(\tau) = O(\delta^2)$; (ii) blue curves correspond to vanishing of the quadratic term  $c_2=0$ in $x(\tau)$, given by  Eq. (\ref{eq:system:sin2}). Note, that under the last condition $c_2=0$ the linear in $\delta$ term may survive ($c_1\neq 0$). Crossing points   of these blue and red set of curves satisfy both of the conditions (i, ii). These points correspond  to synchronization of  qubits excitations with the precision  $n_0(\tau,\delta)\propto (\delta/\Omega)^6$. The third parameter $k_1$ is found from (\ref{eq:system:phi}) where we set $N=3$ and take $k_3, k_5$  in accordance with  blue and red curves crossing points, e.g. that one is  marked as  greed dot. This green point corresponds to  envelope $f(t)$ having smallest maximum value  which provides the effective suppression of the disorder.

\subsection{Synchronization of the qubits excitation vs detuning}

\begin{figure}[!ht]%
\centering
\subfloat[][$N=1$]{%
\includegraphics[width=0.45\linewidth]{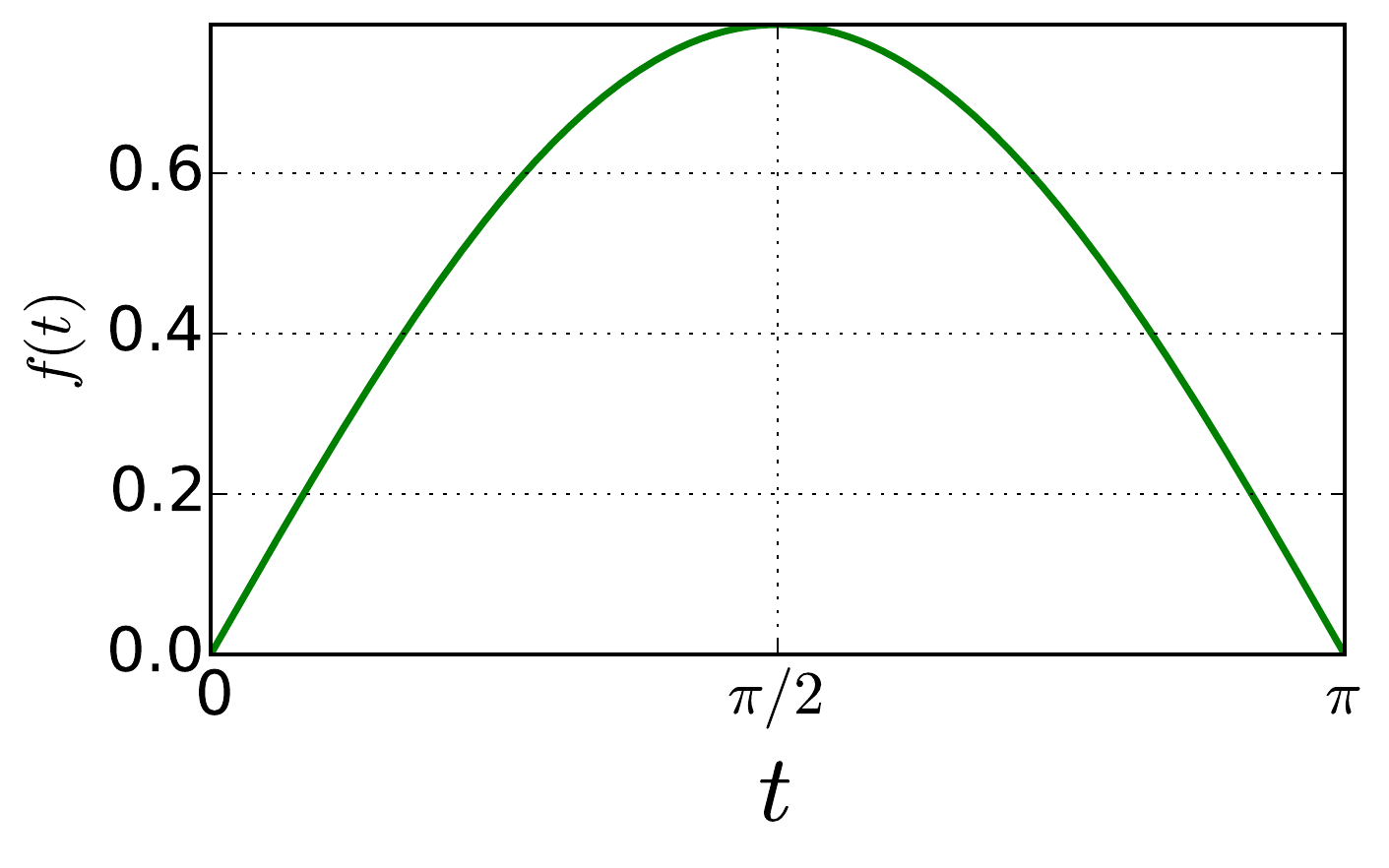}%
\label{fig:forms:1}%
\includegraphics[width=0.45\linewidth]{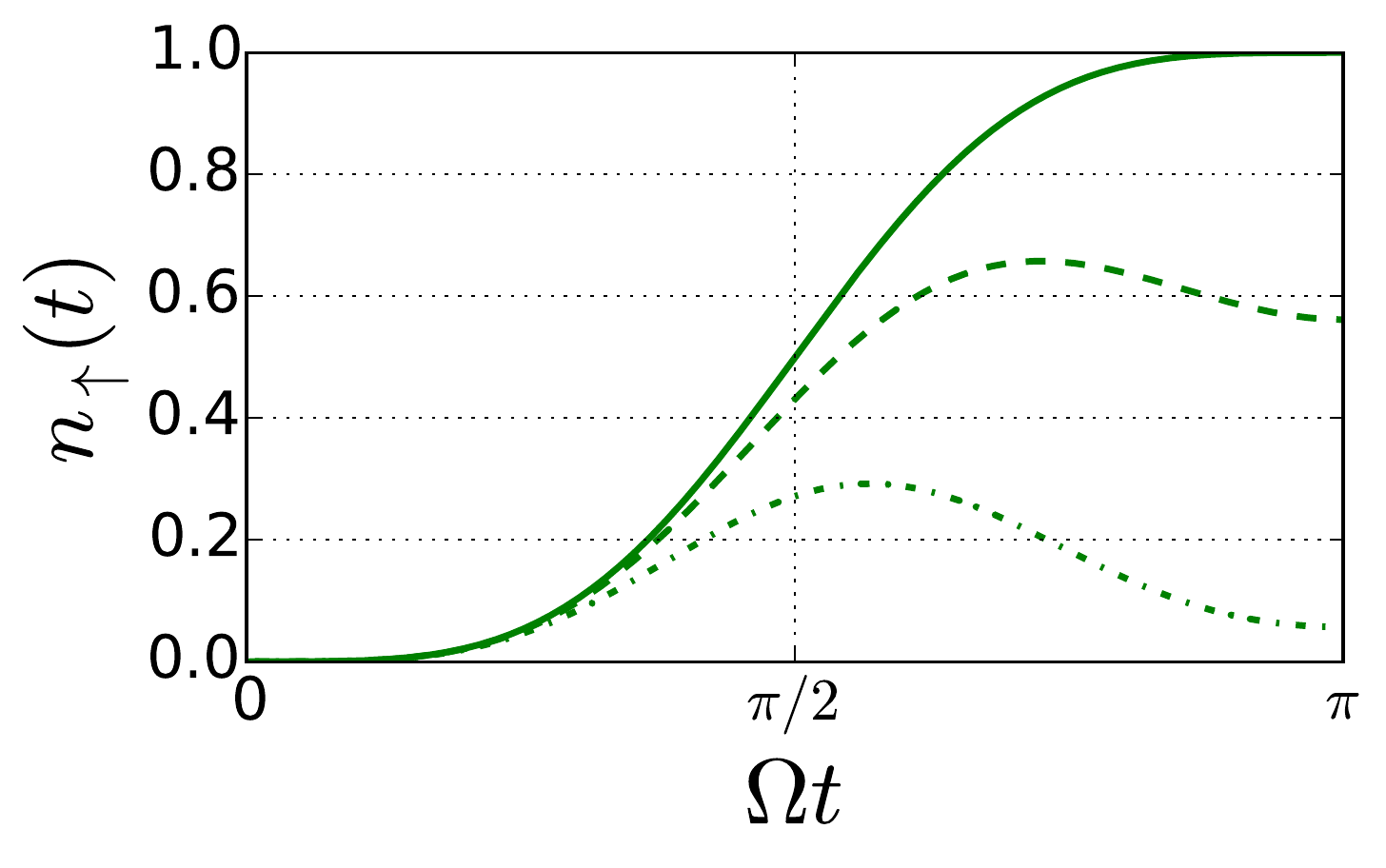}%
\label{fig:forms:1}}
\\
\subfloat[][$N=2$]{\includegraphics[width=0.45\linewidth]{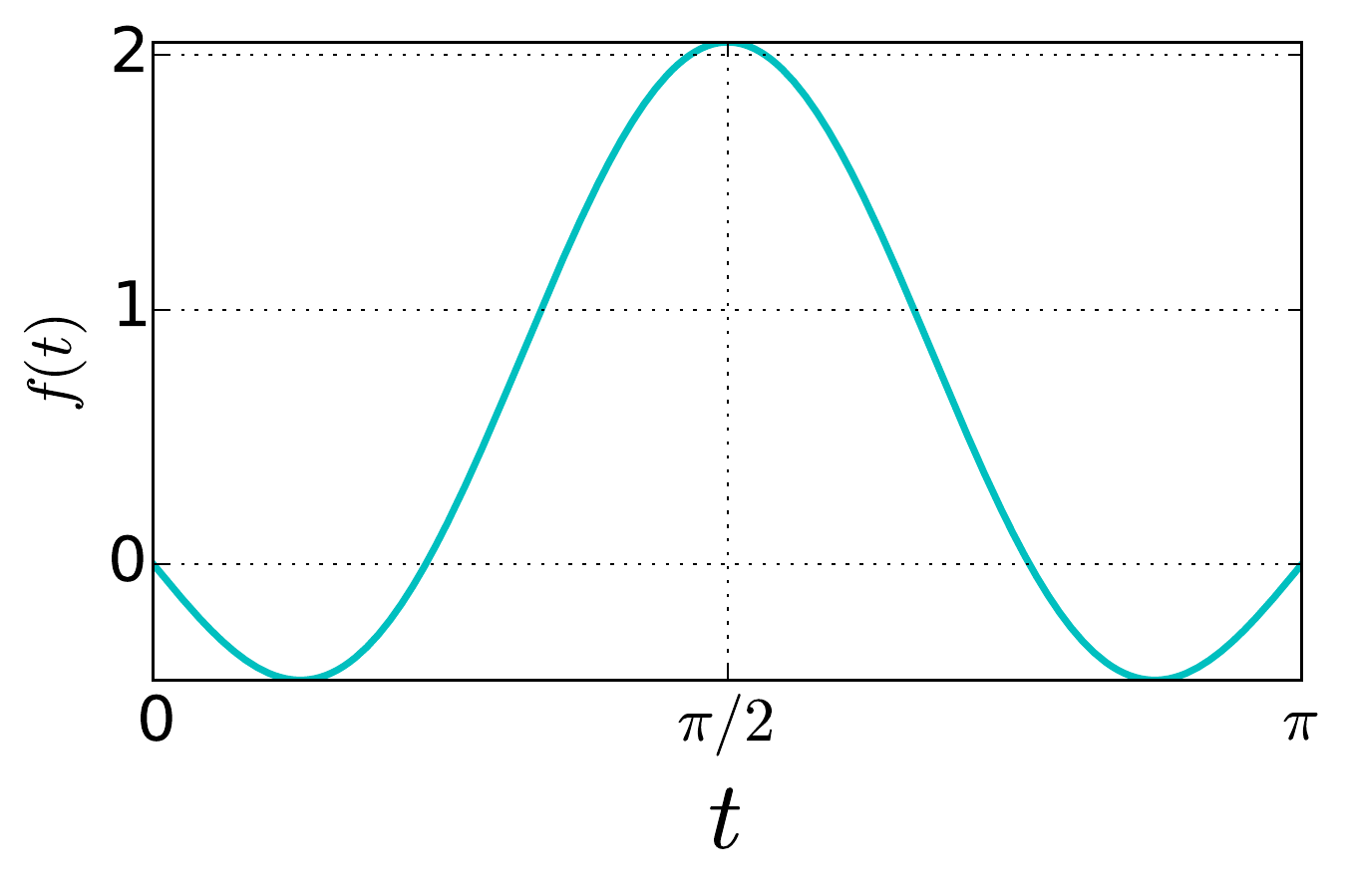}\label{fig:forms:2}
\includegraphics[width=0.45\linewidth]{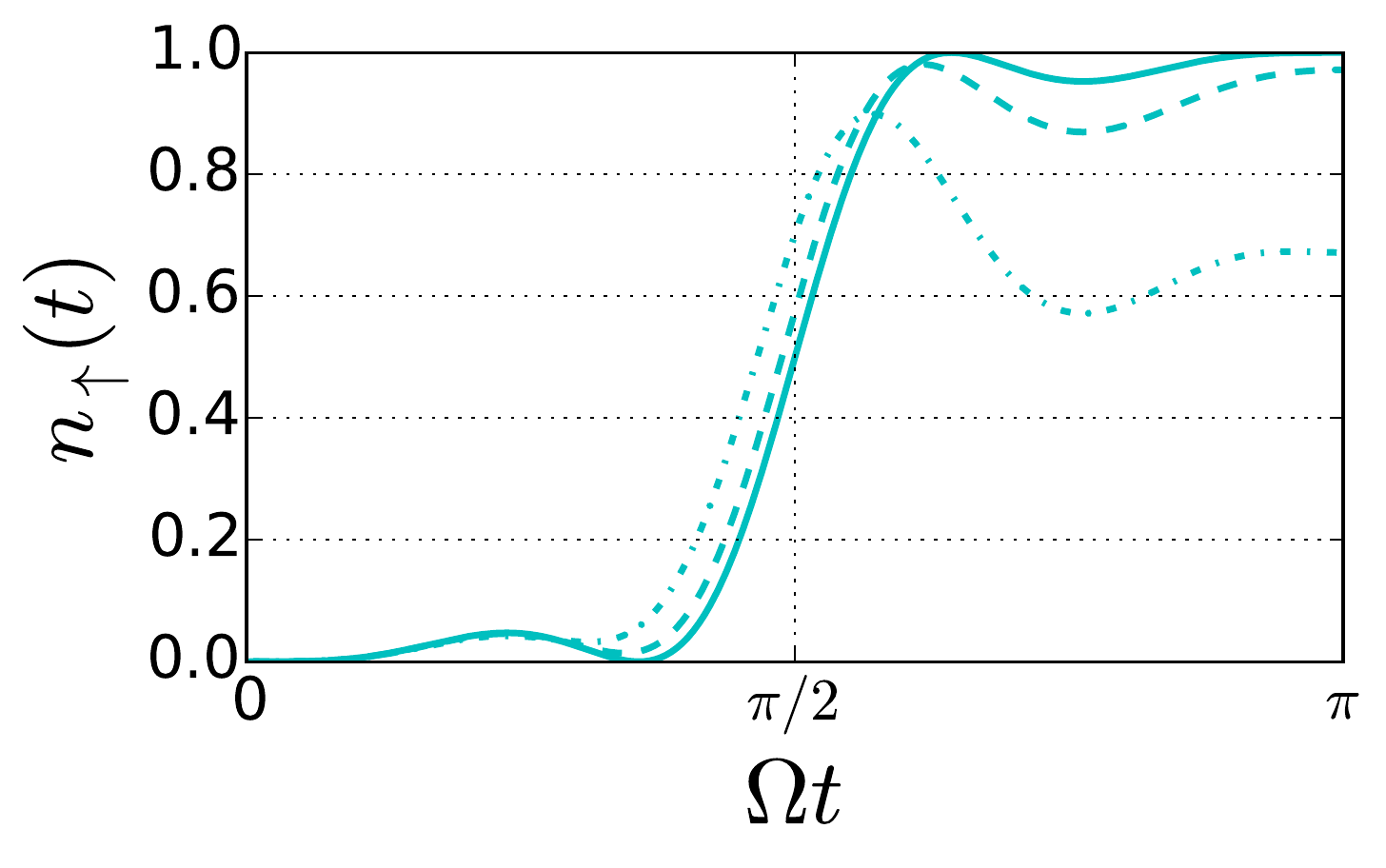}\label{fig:forms:2}}
\\
\subfloat[][$N=3$]{\includegraphics[width=0.45\linewidth]{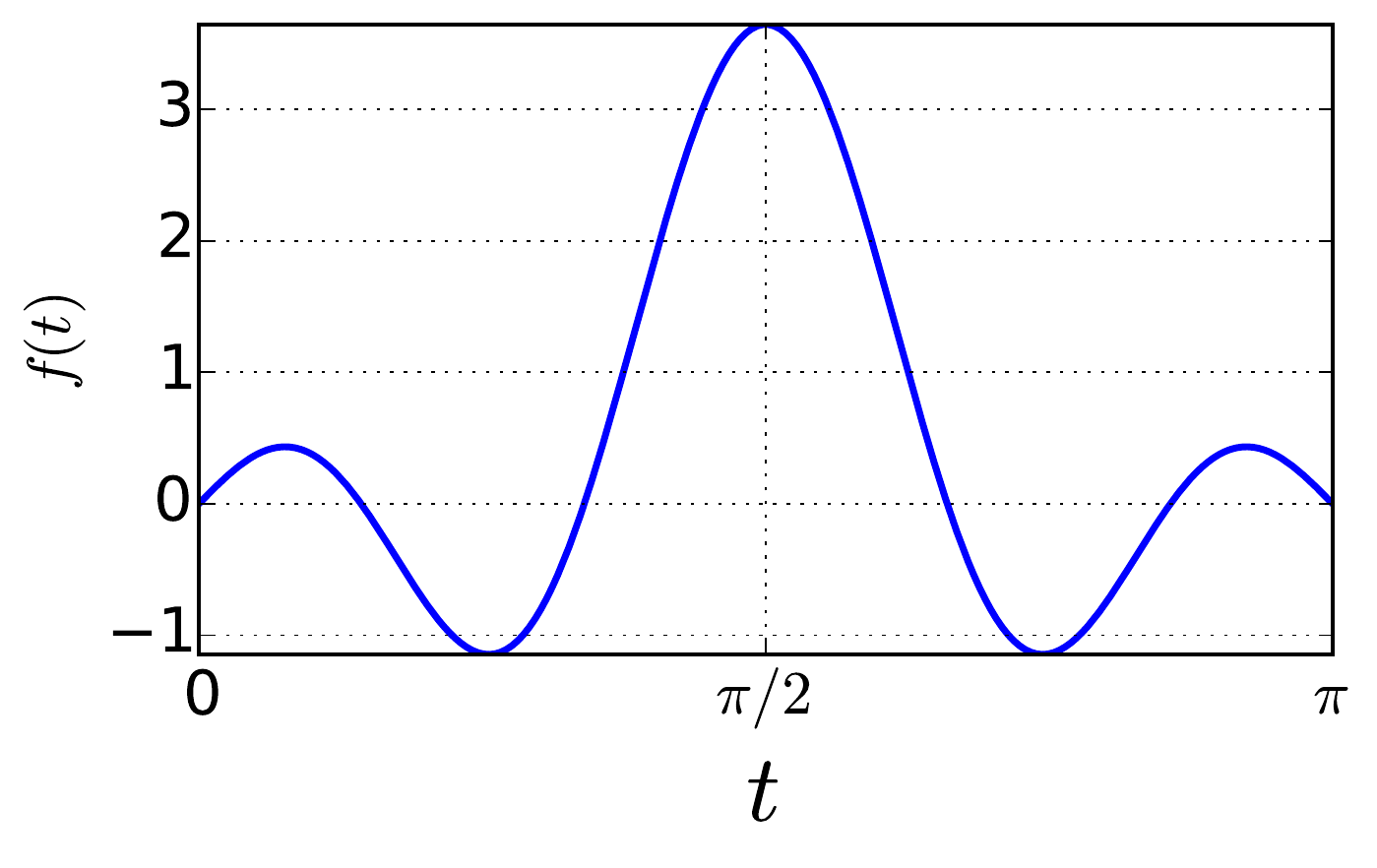}\label{fig:forms:3}
\includegraphics[width=0.45\linewidth]{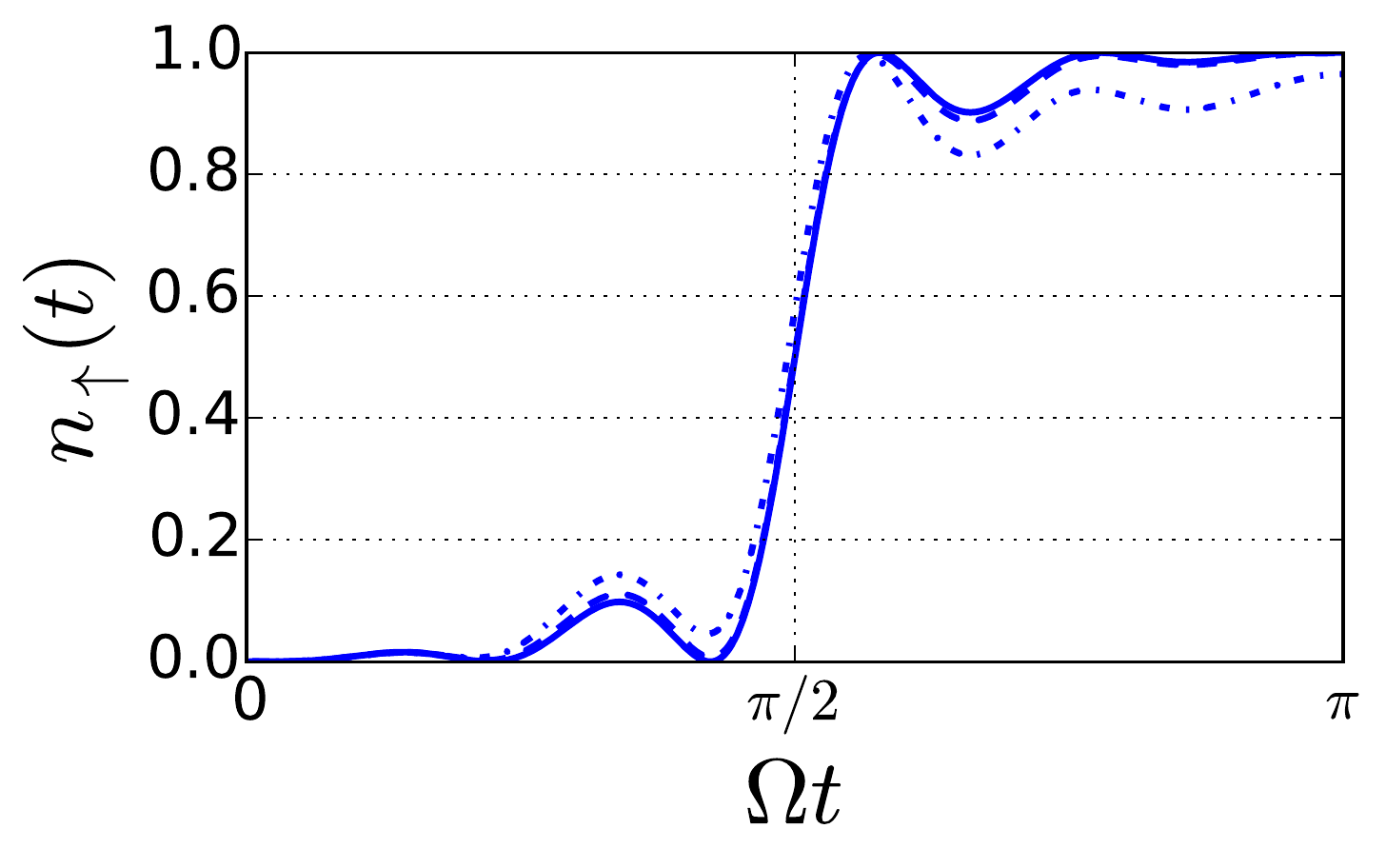}\label{fig:forms:3}}%
\\
\subfloat[][$N=4$]{\includegraphics[width=0.45\linewidth]{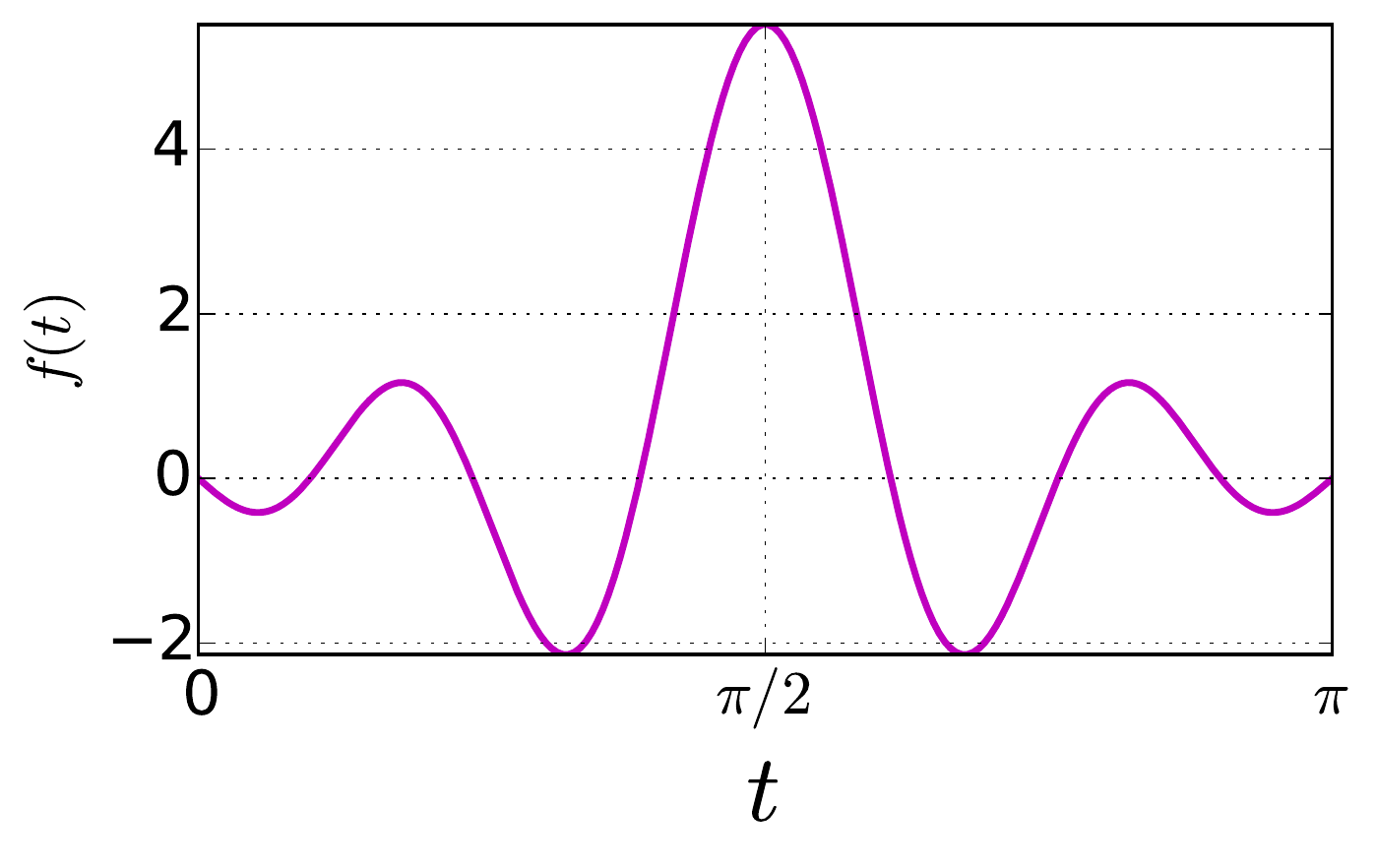}\label{fig:forms:4}
\includegraphics[width=0.45\linewidth]{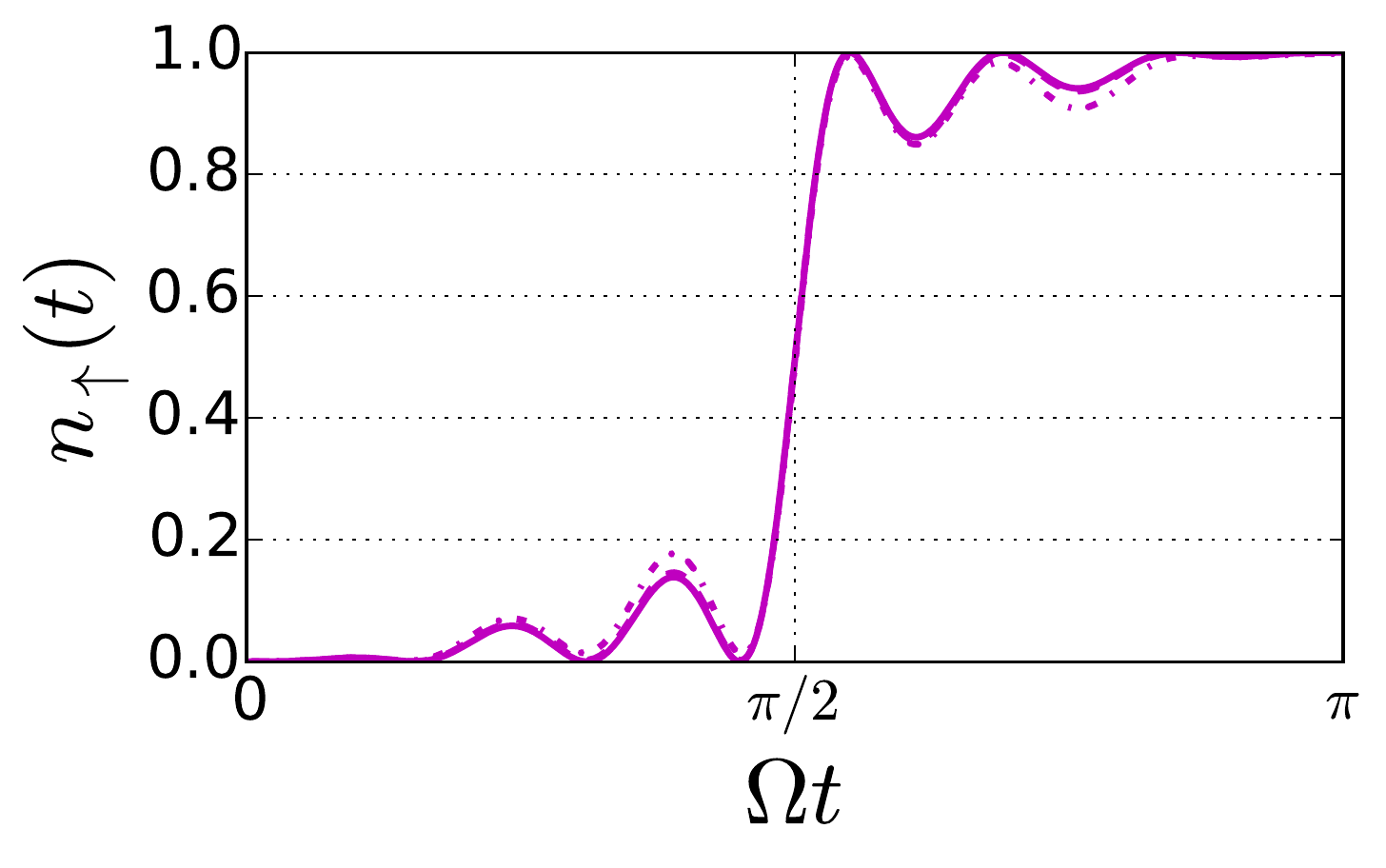}\label{fig:forms:4}}
\caption{The shapes of driving amplitudes (on the left) and time dependence of excited level occupation number $n_\uparrow(t)$ (on the right) plotted for schemes of a different orders. The driving envelope $f(t)$ shown in fig.\ref{fig:forms:1} corresponds to non-optimized $\pi$-pulse and is plotted as a reference.}
\label{fig:profiles}
\end{figure}

 The higher order schemes are build straightforward around the above solution at $N=3$.
 In this section we collect all the results for $N=1,2,3,4$  order schemes  in the driving envelope function (\ref{eq:phisum}). In the left column of the fig.\ref{fig:profiles} we plot the optimized shapes of $\pi$-pulses found within the above perturbative  approach for a given truncation number $N$. In the right column we show plots illustrating time evolution of qubit excitation dynamics  $n_{\uparrow}(t)$ within $\pi$-pulse duration time  $0<t<\tau$. Solid curves in the right column in fig.\ref{fig:profiles}   correspond to resonant driving $\delta=0$, while dashed ones are related to the detuning $\delta=\Omega$ and $2\Omega$. The dynamics  starts from the ground state at $t=0$ and grows significantly at the half of the $\pi$-pulse duration time $\tau/2$. The increase  of the $n_\uparrow(t)$ resemble the response to singular driving at $t=\tau/2$, because the limit of $f(t)$ at $N\to\infty$ corresponds to ideal $\pi$-pulse  with $f_{N\to\infty}(t) =\pi\delta(t-\tau/2)$, which is obviously not achievable experimentally. We stress that we work in the regime of  finite $N$ and amplitudes and treat the efficiency of this approach   by means of deviation of  the resulted $n_\uparrow(\tau)$ from unity with respect to non-zero detuning $\delta$.
The last two plots  in fig.\ref{fig:profiles} illustrate  good efficiency of the corresponding  $\pi$-pulse shapes:  at $N=3$ and $4$ we observe that the dashed curves are very close to the solid ones at $t=\tau$.
This means that  the inhomogeneous broadening is  effectively  suppressed in a Rabi frequency range $\propto \Omega$ and this technique
offer the  synchronization of qubit ensemble starting even from $N=3$.

\begin{figure}[!ht]
  \includegraphics[width=\linewidth]{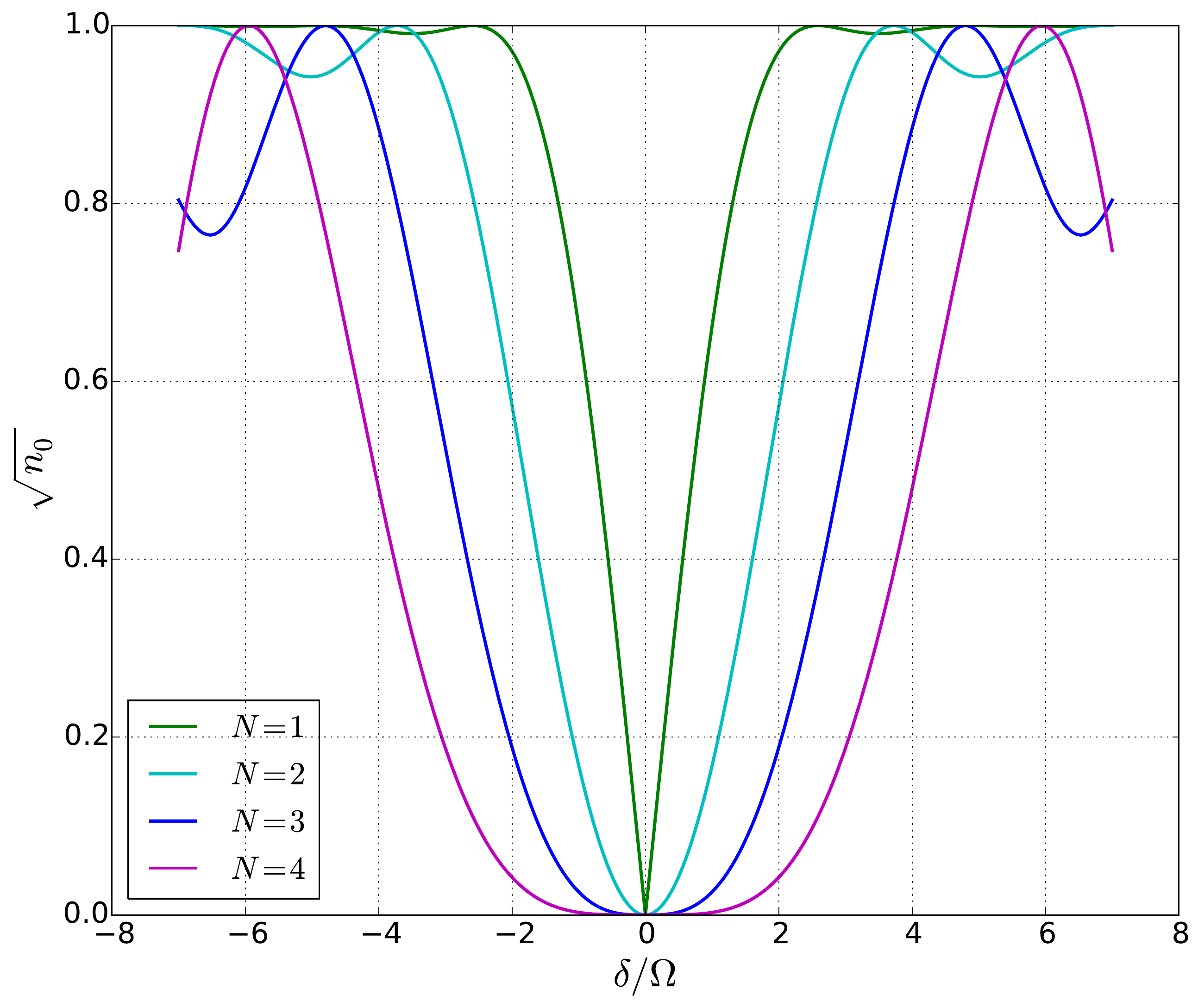}
  \caption{Dependence of ground state amplitude after optimized $\pi$-pulse as a function of detuning in schemes of different orders $N$.}
  \label{fig:gsa}
\end{figure}

In the fig.\ref{fig:gsa} we plot numerical results for the dependencies of ground state amplitude absolute values $\sqrt{n_0(\tau) }$ after  $\pi$-pulse  as a function of detuning $\delta$ associated with a spectral broadening. It can be seen that the increase of  the scheme order  $N$ results in flattening of the curves for $\sqrt{n_0(\tau)}$ around point $\delta=0$. This flattening is  quantitative demonstration of the inhomogeneous broadening suppression.

\begin{figure}[!ht]
  \includegraphics[width=\linewidth]{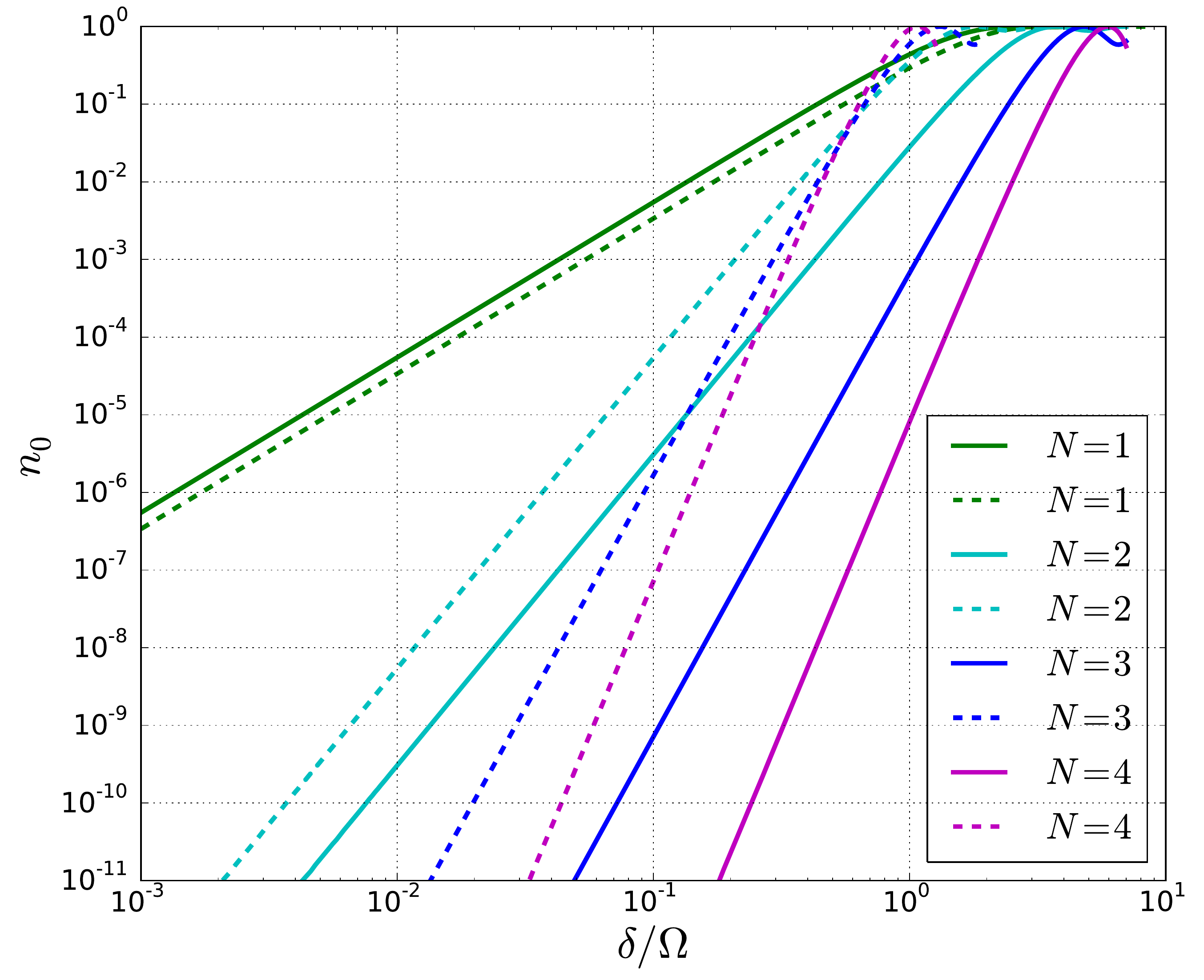}
  \caption{Dependence of  residual  ground state occupation number after optimized $\pi$-pulse on detuning value plotted in double logarithmic scale. Solid lines correspond to optimized $\pi$-pulses of the same duration $\tau=\pi/\Omega$. Dashed lines correspond to normalized envelopes having maximum absolute value equals to unity. In the latter case the $\pi$-pulse time is not fixed and increases with the increase of scheme order.}
  \label{fig:gsa:loglog}
\end{figure}

The fig.\ref{fig:gsa:loglog} is plotted in double logarithmic scale illustrates the precision $ n_0(\tau) $ of the  $\pi$-pulse technique proposed. This figure allows one to estimate the residual value of ground state amplitude for the given scheme order.  For instance,  for $N=3$ the optimized $\pi$-pulse allows us to achieve the probability of qubit ensemble excitation up to $n_\uparrow \approx 1-10^{-3}$ at detuning  values up to $\Omega$.

The driving amplitude is limited in   real experiment. To illustrate the effect of this limitation we plot the residual ground state amplitude $x(\tau)$ for the pulse of the same shape as mentioned above, but at different $\tau$ so that maximum value of the corresponding envelope does not exceed $\Omega$. These curves are plotted as dashed lines in fig.\ref{fig:gsa:loglog}.

\begin{figure}[!ht]
  \includegraphics[width=\linewidth]{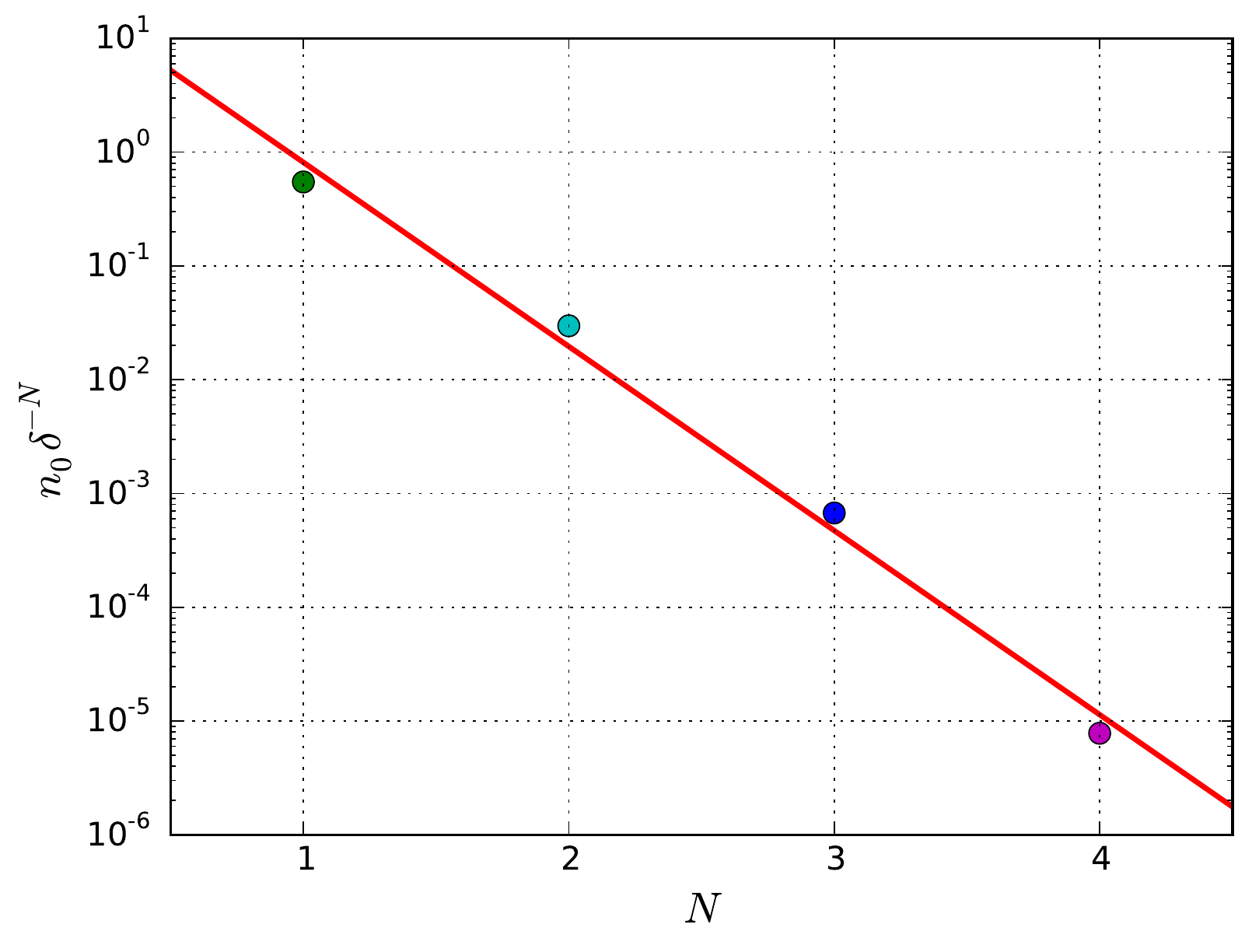}
  \caption{Ground state occupation normalized by appropriate power of detuning after optimized $\pi$-pulse as function of scheme order $N$ involved in $f(t)$.}\label{fig:gsa:loglogN}
\end{figure}

From  the curves for $\sqrt{n_0(\tau)}$ shown  in fig.\ref{fig:gsa:loglog}  we extract the coefficients of power-law dependencies of residual ground state amplitude   for a given $N$. As one can see, the dimensionless combination $n_0(\delta)(\delta/\Omega)^{-N}$ does not depend on $\delta$. Dependence on $N$  of logarithmically scaled value of this combination can be fitted by a straight line, as shown in fig.\ref{fig:gsa:loglogN}
\begin{equation}
  n_0(\delta)=1-n_\uparrow \approx
  34
  \left(
  0.16
  \frac{\delta}{\Omega}\right)^{2N} \label{eq:n0}
\end{equation}
Eq. (\ref{eq:n0}) is one of central results which show quantitative dependence of the precision on the order $N$ and detuning.
The small scaling factor
$0.16$
for $\delta/\Omega$ shows that this technique based on sine representation (\ref{eq:phisum}) could provide the synchronization of ensemble even if the driving signal amplitude is less than a broadening by one order of the magnitude.


\section{Discussion}
The efficient excitation of NV-centers in diamond reported in Ref. \cite{nv-centers}  were achieved through the sequences of rectangular  pulses with periodical switching of the amplitude sign. This technique demonstrates possibility of excitation of strongly off-resonant qubits by a weak driving signal.  We studied an opposite regime  when  the single non-rectangular $\pi$-pulse  effectively  suppresses the disorder. The mechanism we are addressed to   is related to synchronous excitation dynamics of two level system under a particular non-rectangular envelope shape  $f(t)$ of the $\pi$-pulse. We considered non-rectangular smooth shape  of the $\pi$-pulse given by external electromagnetic  driving   $F(0<t<\tau)=f(t)e^{-i\omega t} $ with the envelope representable as the sum of $N$ sine functions $f(t)= \Omega\sum^N k_{2n+1}\sin (2n+1)\pi t/(2\tau)$ where the amplitude and pulse duration are locked with each other $\tau=\pi/\Omega$. The  off-resonant response of a qubit to a non-rectangular signal can not be calculated exactly and we found the perturbative solution. Within this solution we proposed the method based on optimization of the set of $N$ parameters $k_{2n+1}$ which provide synchronous  excitation of the off-resonant qubits. Note, that this optimization is not direct expansion in sine basis of ideal $\pi$-pulse in delta-functional form. The precision of this method, expressed in terms of qubit excitation number $n_\uparrow$, is controlled by the order of the scheme and  proportional to $(\delta/\Omega)^{2N}$. This scheme is efficient for the qubit  energies falling into tunable spectral range estimated as the  driving amplitude strength $\Omega$. Within our solution  we demonstrated  that the $\pi$-pulse formed by $N=4$ sine functions
shows simultaneous excitation of qubits with the probability up to $n_\uparrow \approx 1-10^{-5}$ for qubit frequencies ranging in $\approx\omega\pm \Omega$.

The sine expansion we used in this approach was based on the experimental  requirement of continuity of $f(t)$ at initial and finite moments of time. Optimal envelope shape $f(t)$ can also be found in another basis, say cosine series or rectangular-based blocks. Our calculations shows that in these cases the results will be qualitatively the same as described above. Thus, the proposed method is quite general and can be tuned to meet requirement and restrictions of a particular experiment.

To conclude, we propose the model of smooth shaped single $\pi$-pulse which can be applied to realistic disordered qubit ensembles coupled to a  transmission line. Such a $\pi$-pulse provides an effective suppression of the inhomogeneous broadening and can be used as qubit synchronization technique.
We expect that our findings could serve as a complementary  methods to those reported in Ref.\cite{nv-centers}, where the sequences of rectangular pulses were used to increase the  efficiency in the excitation of    qubits within   certain frequency range. We also mention that similar technique can be effectively applied to create, say, $\pi/2$-pulse to prepare entangled states of the inhomogeneously broadened qubits.

\section{Acknowledgments}
Authors thank Alexey V. Ustinov and Walter V. Pogosov for discussions.

\end{document}